\documentclass[journal]{IEEEtran}
\usepackage{ifpdf}
\usepackage{cite}
\ifCLASSINFOpdf
   \usepackage[pdftex]{graphicx}
\else
\fi
\usepackage{subcaption}
\usepackage{multirow, makecell}
\usepackage{textcomp}
\newcommand{\sfdagger}{{\sffamily\textdagger}}

\usepackage{amsmath,soul}
\usepackage{amsfonts,amsthm}
\usepackage{amssymb}
\usepackage{amsthm}

\usepackage{bm}
\usepackage{algorithmicx}
\usepackage{algpseudocode}
\usepackage{enumitem}
\usepackage[Algorithm,ruled]{algorithm}

\usepackage{stfloats}

\usepackage{kotex}
\usepackage{tabu}

\begin{document}
\title{Dynamic and Super-Personalized Media Ecosystem Driven by Generative AI:\\ Unpredictable Plays Never Repeating The Same}

\author{Sungjun Ahn, \emph{Member}, \emph{IEEE}, Hyun-Jeong Yim, \emph{Member}, \emph{IEEE}, Youngwan Lee, and Sung-Ik Park, \emph{Fellow}, \emph{IEEE}
        % <-this % stops a space
%\thanks{This work was supported by Institute of Information $\&$ communications Technology Planning \& Evaluation (IITP) grant funded by the Korea government [MSIT, Development of Receiver Chip for ATSC 3.0 Mobile Broadcast] under Grant RS-2023-00224660. \textit{(Corresponding author: Sungjun Ahn.)}}
\thanks{Sungjun Ahn, Hyun-Jeong Yim, and Sung-Ik Park are with the Media Research Division, Electronics and Telecommunications Research Institute (ETRI), 218 Gajeong-ro, Yuseong-gu, Daejeon, 34129 South Korea (e-mail: $\{$sjahn, hjyim, psi76$\}$@etri.re.kr)}
\thanks{Youngwan Lee is with the Intelligence Information Research Division, ETRI, 218 Gajeong-ro, Yuseong-gu, Daejeon, 34129 South Korea (e-mail: yw.lee@etri.ac.kr)}}

\maketitle

\begin{abstract}
This paper introduces a media service model that exploits artificial intelligence (AI) video generators at the receive end.
This proposal deviates from the traditional multimedia ecosystem, completely relying on in-house production, by shifting part of the content creation onto the receiver.
We bring a semantic process into the framework, allowing the distribution network to provide service elements that prompt the content generator, rather than distributing encoded data of fully finished programs.
The service elements include fine-tailored text descriptions, lightweight image data of some objects, or application programming interfaces, comprehensively referred to as semantic sources, and the user terminal translates the received semantic data into video frames.
Empowered by the random nature of generative AI, the users could then experience super-personalized services accordingly.
The proposed idea incorporates the situations in which the user receives different service providers' element packages; a sequence of packages over time, or multiple packages at the same time.
Given promised in-context coherence and content integrity, the combinatory dynamics will amplify the service diversity, allowing the users to always chance upon new experiences.
This work particularly aims at short-form videos and advertisements, which the users would easily feel fatigued by seeing the same frame sequence every time.
In those use cases, the content provider's role will be recast as scripting semantic sources, transformed from a thorough producer.
Overall, this work explores a new form of media ecosystem facilitated by receiver-embedded generative models, featuring both random content dynamics and enhanced delivery efficiency simultaneously.
\end{abstract}


% \fi
\IEEEpeerreviewmaketitle

\section{Introduction}
\IEEEPARstart{T}{he} last decade has been an era of mushrooming advances in machine learning (ML).
Accelerating progress in ML has impacted the entire discipline of signal and data processing.
Potential leap in multimedia creation and distribution is accentuated since the field of media processing, e.g., image and audio, has driven the boost in modern ML, and the communications community has been lively at transforming conventional systems into ML-native figures \cite{Dang}, \cite{Quraan}.
For a distribution-side example, definitions of 6G by major players largely share a common vision of integrating native artificial intelligence (AI) capabilities across various network functionalities, yet the details are still contested \cite{Huawei}, \cite{Samsung}.
As widely anticipated, integrating various semantic multimedia processing with such intelligent networks will bring a much-awaited breakthrough reforming the overall landscape.\\
\indent One could tack generative AI (GenAI) and semantic communications (SC) as the most compelling contributors in this context \cite{Bao}-\!\!\cite{Vaswani}.
%A geared coupling between them may reshape the network beyond a deterministic conveyor, an emergent property
It is not an exaggeration to say that GenAI led the latest revolution in ML.
GenAI has moved beyond deterministic task-solving and infused emergent properties into machines, practically enabling them to yield \emph{creativity}.
Natural language processing (NLP) and large language model (LLM) technologies, exemplified by \emph{ChatGPT}, have not only demonstrated their capabilities but also showcased their practical utility through diverse commercial applications \cite{GPT-3}-\!\!\cite{Jiang}.
From the perspective of the media industry, the potential of GenAI is in its general usability, in other words, public accessibility.
LLM and its applications are not limited to simple linguistic and textual goals \cite{BERT}.
Their impressive ability to create imagery and musical craft arts is already enjoyable in public, where \emph{DALL-E} and \emph{AudioCraft} stand as good examples \cite{Ramesh1}-\!\!\cite{Audiocraft}.
Furthermore, GenAIs have also set out into moving scenes, introducing image-to-video (I2V) generators and flexible video re-editors, such as \emph{DeepFake}, which intelligently synthesizes the moving imageries \cite{Zhang_XL}-\!\!\cite{Ni}.\\
\indent All these possibilities fuel SC and push it to practice.
The history of SC has long lasted for decades \cite{Weaver}, marked by ongoing efforts to leverage semantic information in data spaces, including initiatives such as MPEG-7 \cite{Salembier}. 
However, it remains a fact that there has been no notable milestone in this context.
As described in \cite{Bao}, receiver-mounted GenAIs can facilitate SC by producing the complete media content, probably sizable, from received semantic information, so that it reduces network burden.
SC escapes from a rigid framework of traditional communications, which have sought to ace the bit-level accuracy test, namely attempting a sound replay of the original source \cite{Xie}-\!\!\cite{Luo}.
Whenever the symbolic features and intentions are correctly delivered, SC allows enjoying presentation diversity and dynamics since it leaves the appearance of service up to GenAI's random \emph{imagination} \cite{Qin}.
Similarly, the multimedia use case of interest presumes a user-side system fabricating contents from the received data.
We primarily deal with video content, but the output format may include audio, text, or tactile stimuli, basically accessible through a human interface.
Moreover, SC also offers a joy of variation that revitalizes the user experience.\\
\indent This paper focuses on such service content dynamics coming out from the collaboration of GenAI and SC.
We consider the future landscape with GenAIs installed in multimedia receiver devices and present the possible transformation of media networks therein.
Device-mounted GenAI is no longer the future; it has already become a reality.
Smartphone manufacturers, Samsung and Apple, for example, have commenced installing local LLM in their appliances \cite{Smith}, and brisk development in model compression will accelerate such GenAI deployment.
As a sequel, \emph{creativity} in content production could partly migrate to end devices.
We suggest this migration as a part of the semantic-native direction envisioned in 6G broadcast intelligence.\\
\indent Specifically, this paper describes a new transformation of delivery systems in a high-level view, focusing on the scenarios where the \emph{things}-to-video generation takes place at the user device.
For short-length media scenarios, in which the proposed system would be more viable, broadcasters would deal with \emph{semantic information} only.
This will rather ease the broadcasters' responsibility related to in-house full production. 
This paper defines the viable form of a \emph{semantic source} enabling the proposed system, and further discusses additional concerns and challenges.
As mentioned, the proposed system utilizes random generation that leads us to a super-personalized experience.
That is, users are allowed to experience a different service each time, even while receiving identical data.
Amid the growing trend toward personalized media, this idea is poised to meet contemporary demands while still aligning with the legacy of linear broadcasting networks.\\
\begin{figure*}[!t]
\centering
\vspace{-0.1cm}
\includegraphics[width = 1.5\columnwidth]{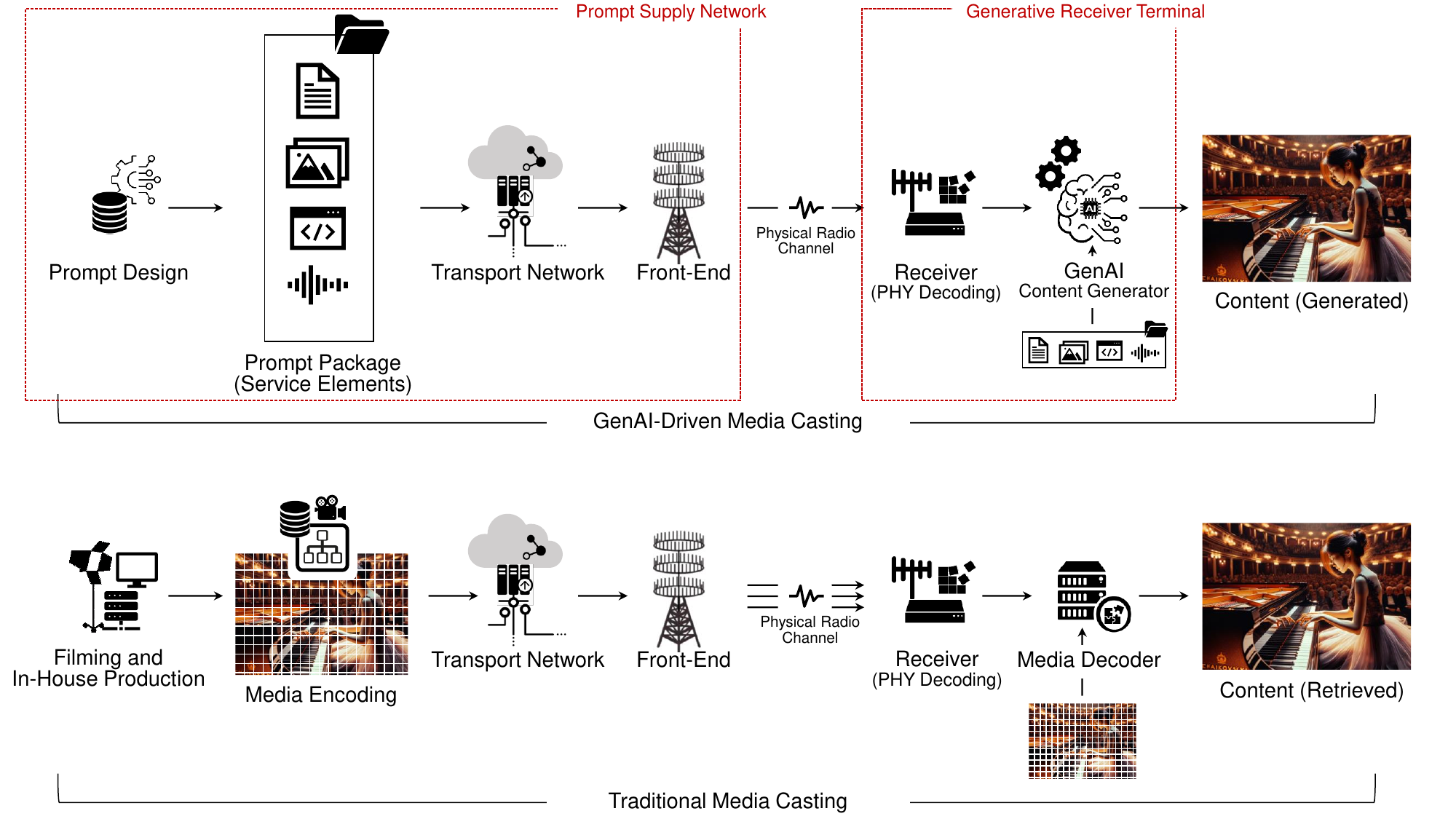}
\vspace{-0cm}
\caption{Proposed GenAI-driven media casting vs. traditional media casting.}\label{PDM_illust}
\vspace{-0.4cm}
\end{figure*}
\indent The definition of semantic sources is explored primarily by recognizing them as prompts but is not confined to text alone. 
It is comprehensively described to encompass a broader scope that includes operational paradigms, application programming interfaces (APIs), and engines.
Continuing from the standardization efforts and discussions on previous semantic media descriptions, this discussion also delves into issues associated with establishing a standardized foundation.
Additionally, we also propose multi-prompt operations extending the proposed system.
The proposed idea incorporates the situations in which the user receives different service providers' element packages; a sequence of packages over time, or multiple packages at the same time.
Given promised in-context coherence and content integrity, the combinatory dynamics will amplify the service diversity, allowing the users to always chance upon new experiences.
This work particularly aims at short-form videos and advertisements, which the users would easily feel fatigued by seeing the same frame sequence every time.
Overall, this work explores a new form of media ecosystem facilitated by receiver-embedded generative models, featuring both random content dynamics and enhanced delivery efficiency simultaneously.\\
\indent The rest of the paper is organized as follows.
Section II provides a review of the related previous works.
The proposed system is elaborated in Section III, first describing the architecture with single-source operation and then extending it into multi-source scenarios.
Possible use cases are discussed in Section IV, and Section V concludes the paper with remarks.
%The magnificence of this innovation lies in its ability to embrace abstract system operations, a departure from traditional approaches that relied on intricate, but tangible, models.
%Instead of struggling with complicated derivations for mathematically provable system functionality, neural network-driven ML approaches have allowed system modules to remain as blackboxes.
%In that point, heuristic demonstrations of working products\footnote{e.g., \emph{AlphaFold}, \emph{AlphaGo}, and \emph{WaveNet} built by DeepMind Ltd.} have first driven the explosive growth of ML in the 2010s.\\
%\indent Rendering abstract features accessible has endowed ML with versatility in the realm of human interaction and experience.
%reproducing sound image text semantic

\section{Advances in Generative Models and Semantic Media Description}
This section reviews the related legacies on which the proposed system is grounded.
Specifically, the parallel progress in generative models, prompt engineering, and media semantics description are summarized.
This brief overview also discovers possible conjunctions among those disciplines, so providing a rational background for the system described in the next section.
\subsection{Text-to-Image (T2I), I2V, and Text-to-Video (T2V) Generation}
\subsubsection{T2I}
Primarily speaking, generative image synthesis is where the latest deep learning models have remarkably augmented the capacity.
T2I synthesis problem has been tackled accordingly, where \cite{Mansimov} inspired many to construct graphics from unstructured text.
The following work, \emph{StackGAN} \cite{StackGAN}, designed a generative adversarial network (GAN)-based system providing $256 \!\times\! 256$ resolution.\\
\indent \emph{DALL-E} \cite{Ramesh1} was a pivotal work that brought zero-shot image generation to the table, which first relied on discrete variational autoencoder (dVAE) architecture.
This means that the system began to compose new imageries that never existed in the dataset.
Such a creative synthesis ability has been further transplanted into diffusion models \cite{DDIM}, which proved qualified fidelity in image generation tasks \cite{Dhariwal}.\\
\indent The first diffusion-based descendant was \emph{GLIDE} \cite{GLIDE}.
This text-guided model well-demonstrated inpainting and creative synthesis unseen in class-guided diffusion \cite{Dhariwal}, but still suffered from the limited performance of its inherent text Transformer \cite{Vaswani}.
Another descendant, \emph{Imagen} \cite{Imagen}, solved this problem by acquiring text embedding from pre-trained LLMs, e.g., T5 \cite{T5}, CLIP \cite{CLIP}, BERT \cite{BERT}, GPT \cite{GPT-3}.
\emph{Imagen} also appended a super-resolution (SR) \cite{SR} function upscaling the generated image into $1024 \!\times\! 1024$ scale, thereby improving aesthetic and resolution qualities at the same time.\\
\indent In parallel, \emph{DALLE-2} \cite{Ramesh2}, leveraging CLIP latent, showcased impressive image manipulation ability that can interpolate the image figures between different intention vectors.
This manipulation functionality, powered by CLIP's cross-modal alignment, made figures controllable within a spectrum between multiple visual concepts; which came special because it enlightened some directions to provoke dynamics for varying image frames and maintain context coherence upon heterogeneous inputs.\\
\indent T2I systems were subsequently extended to embrace multi-modality.
First, \emph{Stable Diffusion (SD)} \cite{StabDiff} diversified available input formats (e.g., layout) while the system still had to receive only one modality at once.
A mesh-up prompting among multiple input modalities was allowed by \emph{GLIGEN} \cite{GLIGEN}, thereby clueing potentials to augment further various modalities.
\subsubsection{I2V and T2V}
Such advancements in image synthesis propelled generative models to branch out into video generation tasks.
\cite{Ho_VDM} lied as the pioneering work for video diffusion models (VDMs), and its cascaded extension \emph{Imagen Video} \cite{Imagen_Video} followed by improving visual quality.
The spatiotemporal U-Nets in \cite{Ho_VDM} and \emph{Imagen Video} steered the research mainstream to VDM from other approaches \cite{CogVideo}-\!\!\cite{GODIVA}, but their architectures still required massive text-video pairs to learn.\\
\indent This was seen as a critical drawback because T2V studies had been troubled by the scarcity of large datasets with high-quality text-video pairs \cite{Singer}.
\emph{Make-A-Video} \cite{Singer} tried to solve this problem by breaking the dependency on text-video pairs, particularly allowing the transfer of T2I knowledge to T2V generation.
Though this approach was already tried before in \cite{CogVideo}, \cite{Singer} improved the performance by fine-tuning the T2I subsystem, whereas \cite{CogVideo} took the pre-trained T2I part as frozen.\\
\indent \textbf{Remark 1 (Towards multi-modality):} Deviating from the cascaded VDM architectures of \emph{Imagen} and \emph{Make-A-Video}, more recent works applied SD, banking on its multi-modal capabilities \cite{LVDM}-\!\!\cite{Magic_Video}.
Moving further, \emph{I2VGen-XL} \cite{Zhang_XL}, \emph{Pika Labs} \cite{Pika}, and \emph{Gen-2} \cite{Gen-2} realized I2V platforms that can be conditioned by image reference and text guidance at the same time.
Such progress suggests a way forward for automated media content generation to embrace a broader range of interface modalities.
This vision partly serves as the motif of our study, reaching an idea of structured wide-modal interfaces to feed generative composers efficiently.\\
\indent Still, video composition tasks remain as challenging.
Human perception is highly sensitive to spatial and motional incoherence, or discontinuity of objects as well, consequently making film art difficult to satisfy the marginal level of completion.
Nonetheless, recent efforts such as \emph{VideoCrafter1} \cite{Chen} and \emph{Emu Video} \cite{Girdhar} brought substantial refinement, and incremental improvement will continue until generative models beat man-made production. \hfill$\qed$\\
\indent %On the other dimension, narrative is another vital requisite for video content.
\textbf{Remark 2 (Narrative generation):} On the narrative front, drastically enhanced LLMs recently empowered GenAIs to be capable of composing short pieces.
The studies on human-AI co-creation platforms, accordingly, introduced an idea to integrate this writing competence of LLM with superficial video generation models, showing promise for automating the entire production of cinematic video content.
For example, \emph{ReelFramer} \cite{ReelFramer} ventured entrusting generative models also with scripting and storyboarding processes in addition to T2V synthesis.
Specifically, clipped news information was converted into short colloquial videos, i.e., news reels, while only three options of narrative framing were available: Expository dialog, reenactment, and comedic analogy.
%This platform could entrust generative models with scripting and storyboarding processes as well.
While narrative consistency remains a challenge and requires human guidance yet, the near future is not perceived as prohibitive for GenAI to produce cinematic videos with artistic drama spontaneously, at least within limited conditions. \hfill$\qed$

\subsubsection{Things-to-Text Feature Extraction}
%Before the efforts on T2I, I2V, and T2V, to extract feature translating 
Conversely, feature extraction from video and image has been extensively tackled before the efforts on T2I, I2V, and T2V \cite{Mou}.
Various studies led AI agents to recognize events, actions, objects, and many features in the visual data, and such techniques were successfully applied into practice.\\
\indent Note that the video conference example in \cite{Bao} first extracts key features from the original video source and only sends differentials of the feature information, so that it reduces the amount of transmission data.
We can remove the feature extraction part and replace the intended feature information with the prescribed semantic descriptions.
For such definitions on semantics, we can refer to the structured profiling previously envisioned in early media protocols, which will be discussed in Section II-C below.
\subsection{Prompt Engineering}
In fact, generative models presented unguaranteed performance depending on prompt commands.
Since generative models became compatible with natural languages, the term \emph{prompt} gained a special meaning in the domain of GenAI:
The input instructions to provoke response back from generative models.
As departed from true-deterministic language and formulaic operations, GenAI systems now have to deal with intensified semantic ambiguity, which hinders promising response relevance consistently.
This problem induced the advent of prompt design technologies, namely, \emph{prompt engineering} \cite{Liu}.\\
\indent Prompt engineering, in essence, has been practice-driven because each instantaneous move of GenAI is beyond our prediction.
Accordingly, third-party individuals have heuristically contributed to this discipline by sharing customized prompt presets \cite{openart}, \cite{flowgpt}.\\
\indent \textbf{Remark 3 (Structured prompt design):} One noticeable approach observed therein is a structured formulation.
This appears interesting when we remember that LLM made the systems open to flexible instructions based on natural language.
According to \cite{Chater}, human intent and imagination are actually vague; \cite{Shafto}, \cite{Seo_TCCN} further say the verbalization process, i.e., representations mapping thoughts to symbols, essentially accompanies uncertainty.
It can be deduced that, if there are specific directions for prompting the response, a structured description format is firmly effective rather than keeping lining in natural sentence patterns.
For media content generation cases that are of our interest, the scenery features could be articulated by specifying the object, style, action, and other details with respect to classified items.\\
\indent We emphasize that such concise prompting can efficiently spark an AI-driven media system. 
We envision a structured template of prompts to ensure product quality.
This idea is motivated by many previous efforts drawn in semantic media standards, which are addressed in Section II-C. 
We can further build a compressive framework dedicated to our purpose. \hfill$\qed$
%Plus, lately emerged small language model (SLM) and prompt augmentation/expansion.
\\
\indent \textbf{Remark 4 (Inclusion of chain-of-thought):} Another notable discovery in prompt engineering is chain-of-thought (CoT) prompting \cite{Wei_CoT}.
Breaking the instruction into stepwise pieces is proven to achieve more relevant inference, in both empirical \cite{Wei_CoT} and information-theoretic \cite{Jiang} fashion.
The observations in CoT shifted the paradigm from fine-tuned prompting to in-context learning.
The structured \emph{prompt templates} may include \emph{automatic thought sequence}, which will deviate from the previous media description frameworks. \hfill$\qed$
% i.e., inconsistently varying quality depending on input prompt configuration. structured
%unsatisfactory

%The field responsible for composing, standardizing, and creating

%than solely on natural language
\subsection{Semantic Media Protocol}
\begin{figure}[!t]
\centering
\vspace{-0cm}
\includegraphics[width = 0.93\columnwidth]{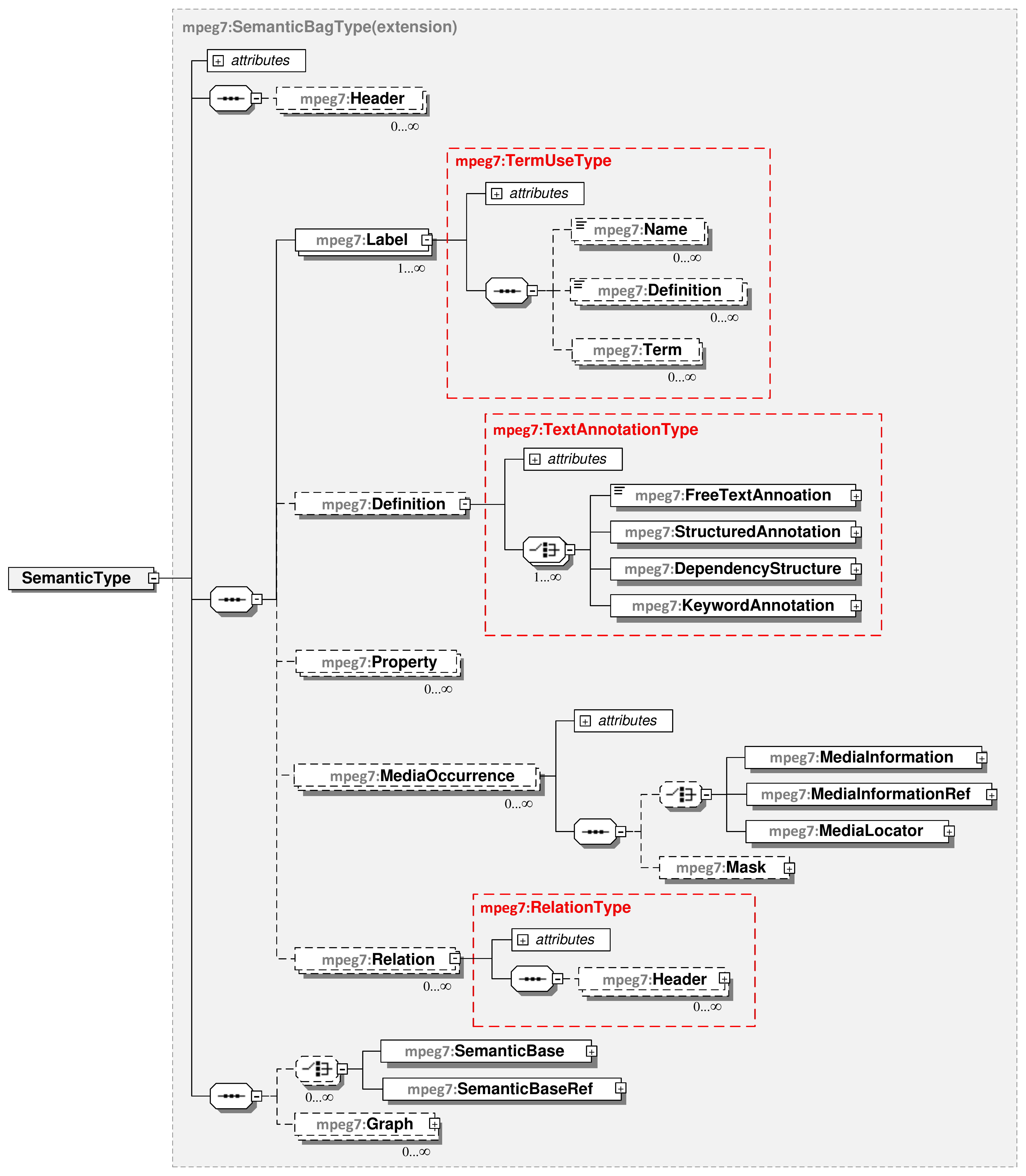}
\vspace{-0cm}
\caption{Structure of MPEG-7 Semantic Description.}\label{xml}
\vspace{-0.5cm}
\end{figure}
\begin{figure}[!t]
\centering
\vspace{-0cm}
\includegraphics[width = 0.45\columnwidth]{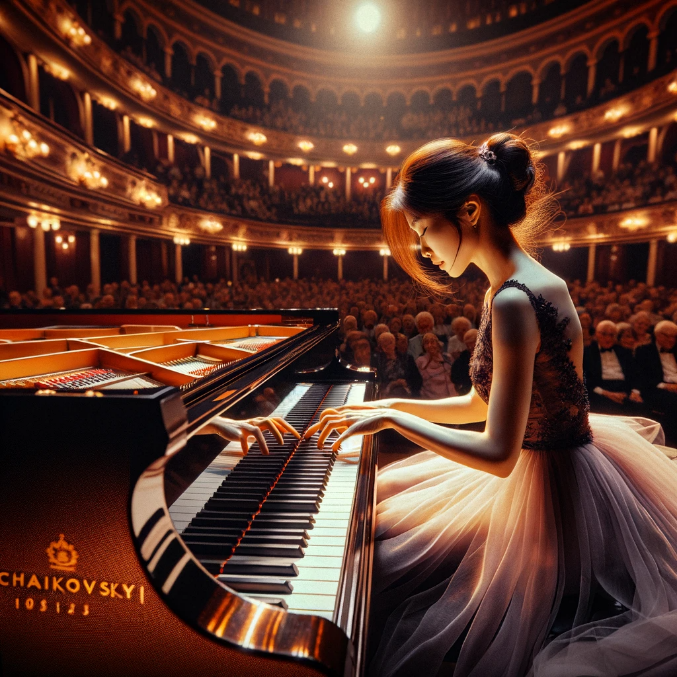}
\vspace{-0cm}
\caption{Image generation example: \emph{DALL-E} prompted based on the description in Config. \ref{MPEG-7_table}. \small{(Script: ``Yim, a renowned pianist, elegantly playing Tchaikovsky's Waltz of the Flowers on a grand piano at a concert hall. The audience, captivated and immersed, watches her performance in a vibrant atmosphere.'')}\normalsize}\label{pianist}
\vspace{-0.2cm}
\end{figure}

As mentioned, the community once established standard structures for semantic media description in the past.
While early efforts deviated from the task of creating media figures, these previous works are still notable, deemed as the cornerstone offering foundational insights.\\
\indent \emph{MPEG-7} is a distinct example to be highlighted, also known as \emph{ISO/IEC 15938 Multimedia Content Description Interface} standard developed by ISO/IEC Moving Picture Expert Group (MPEG).
MPEG-7 defined various Description Schemes to illustrate the information dwelling in media content, e.g., the structure, components, embedded meaning, etc\footnote{Encompasses toolsets such as Visual and Audio Descriptors, Multimedia Description Schemes, and Content Management and Identification Descriptor.} \cite{Martinez}, \cite{Salembier}.\\
\indent The Semantic Description defined in MPEG-7 Part 8 \cite{MPEG7_Pt8}, particularly, allowed describing the events, objects, place, and time in narrative worlds that constitute the media content.
The structure followed the hierarchical design in Fig. \ref{xml}.
%Abstraction for each instance is enabled as well.
Regarding video content, this attempt faithfully captured the elements of a scene of a play, particularly trying to translate scene-by-scene images into verbally conveyable form.
If footage needs to be described briefly$-$approximate implications and composition$-$but as clearly as possible, MPEG-7 can be an effective tool.\\
\indent Config. \ref{MPEG-7_table} presents an example of the Semantic Description configuration.
Note that every statement therein is for describing the image content \emph{YimPlayingPiano.jpg.}
In other words, every description is for labeling the original source file that already exists.
In fact, from a visual perspective, we can recognize that this example consists of one main character playing the piano in a concert hall, no better than that.
The MPEG-7 description is seen as insufficient for guessing the original footage.\\
\begin{algorithm}[!t]\scriptsize
\floatname{algorithm}{Config.}
\caption{Example of MPEG-7 Semantic Description}\label{MPEG-7_table}
$<$Mpeg7$>$\\
.....\\
$~<$DescriptionUnit$>$\\
$~~<$Semantic$>$
\begin{algorithmic}[H]
\State $<$!--Detailed information about the pianist --$>$
\State $<$SemanticBase xsi:type=``AgentObjectType'' id=``PianistYim''$>$
\State \hspace*{\algorithmicindent}$<$Label$>$
\State \hspace*{\algorithmicindent}\hspace*{\algorithmicindent}$<$Name$>$ Yim $<$/Name$>$
\State \hspace*{\algorithmicindent}$<$/Label$>$
\State \hspace*{\algorithmicindent}$<$Definition$>$
\State \hspace*{\algorithmicindent}\hspace*{\algorithmicindent}$<$StructuredAnnotation$>$
\State \hspace*{\algorithmicindent}\hspace*{\algorithmicindent}\hspace*{\algorithmicindent}$<$Who rdf:about=``\#PianistYim''$>$
\State \hspace*{\algorithmicindent}\hspace*{\algorithmicindent}\hspace*{\algorithmicindent}\hspace*{\algorithmicindent}$<$Name$>$Yim$<$/Name$>$
\State \hspace*{\algorithmicindent}\hspace*{\algorithmicindent}\hspace*{\algorithmicindent}\hspace*{\algorithmicindent}$<$Background$>$A renowned pianist known for her interpretation of 
\State \hspace*{\algorithmicindent}\hspace*{\algorithmicindent}\hspace*{\algorithmicindent}\hspace*{\algorithmicindent}classical pieces.$<$/Background$>$
\State \hspace*{\algorithmicindent}\hspace*{\algorithmicindent}\hspace*{\algorithmicindent}$<$/Who$>$
\State \hspace*{\algorithmicindent}\hspace*{\algorithmicindent}$<$/StructuredAnnotation$>$
\State \hspace*{\algorithmicindent}$<$/Definition$>$
\State \hspace*{\algorithmicindent}$<$Relation type=``identity'' target=``\#PianistYim''/$>$
\State \hspace*{\algorithmicindent}$<$Relation type=``hasPerformed'' target=``\#WaltzOfTheFlowers''/$>$
\State \hspace*{\algorithmicindent}$<$MediaOccurrence$>$
\State \hspace*{\algorithmicindent}\hspace*{\algorithmicindent}$<$MediaLocator$>$
\State \hspace*{\algorithmicindent}\hspace*{\algorithmicindent}\hspace*{\algorithmicindent}$<$MediaUri$>$ YimPlayingPiano.jpg $<$/MediaUri$>$
\State \hspace*{\algorithmicindent}\hspace*{\algorithmicindent}$<$/MediaLocator$>$
\State \hspace*{\algorithmicindent}$<$/MediaOccurrence$>$
\State $<$/SemanticBase$>$

\State $<$!--Detailed information about the concert --$>$
\State $<$SemanticBase xsi:type=``EventObjectType'' id=``Concert''$>$
\State \hspace*{\algorithmicindent}$<$Label$>$
\State \hspace*{\algorithmicindent}\hspace*{\algorithmicindent}$<$Name$>$ Concert $<$/Name$>$
\State \hspace*{\algorithmicindent}$<$/Label$>$
\State \hspace*{\algorithmicindent}$<$Definition$>$
\State \hspace*{\algorithmicindent}\hspace*{\algorithmicindent}$<$FreeTextAnnotation$>$The audience is immersed in the performance of 
\State \hspace*{\algorithmicindent}\hspace*{\algorithmicindent}Tchaikovsky's Waltz of the Flowers by pianist `Yim'. 
\State \hspace*{\algorithmicindent}\hspace*{\algorithmicindent}$<$/FreeTextAnnotation$>$
\State \hspace*{\algorithmicindent}$<$/Definition$>$
\State \hspace*{\algorithmicindent}$<$Relation type=``performedBy'' target=``\#PianistYim''/$>$
\State \hspace*{\algorithmicindent}$<$Relation type=``setting'' target=``\#ConcertHall''/$>$
\State $<$/SemanticBase$>$

\State $<$!--Detailed information about the concert hall --$>$
\State $<$SemanticBase xsi:type=``PlaceObjectType'' id=``ConcertHall''$>$
\State \hspace*{\algorithmicindent}$<$Label$>$
\State \hspace*{\algorithmicindent}\hspace*{\algorithmicindent}$<$Name$>$ Concert Hall $<$/Name$>$
\State \hspace*{\algorithmicindent}$<$/Label$>$
\State $<$/SemanticBase$>$
\end{algorithmic}
$~~<$/Semantic$>$\\
$~<$/DescriptionUnit$>$\\
.....\\
$<$/Mpeg7$>$
\end{algorithm}
\indent Nonetheless, we still find the MPEG-7 description acceptable for prompting generative models.
Fig. \ref{pianist} demonstrates a T2I generation prompted by information in Config. \ref{MPEG-7_table}.
Owing to the capacity of the generative model to cope with ambiguity, Fig. \ref{pianist} shows realistic rendering by arbitrarily filling out the parts untold.
Although the prompt script is transformed from the raw template in Config. \ref{MPEG-7_table}, this example shows that the current system works with simple adjustments for interface compatibility.
As a corollary, this trial directs to the validity in video clip cases also. 
Based on efficient graph-based descriptions, MPEG-7 accommodates the relationships among the objects, which will further enrich GenAI's comprehension.\\
\indent However, the existing description tools face challenges because creating content was not an option in their inception.
The existing description tools stem from the Semantic Web, a conception for flexible archives enabling semantic and context-aware connections in data.
%The term \emph{semantic media} first emerged from this context to incorporate the media content domain, i.e., to imbue semantic information into media content, expecting future use in ontology fashion or else.
This means that the existing frameworks were developed for searching specific scenes or content by tracking descriptive \emph{tags}, rather than for directly manipulating the content using semantic data.
Thus, the Semantic Description remains just a labeling, distinguishing individual pieces.
In contrast, video composition demands the assurance of continuity, a more intricate task.
Furthermore, annotations in MPEG-7 Semantic Description are neither obligatory nor consistently defined, often characterized by high arbitrariness.\\
\indent The legacy frameworks defy sophisticated coordination and flexible service.
MPEG-7 lacks abilities to specify the features within the picture domain, i.e., it is impossible to map the semantics to pixels accurately.
Accurate layout definition is not allowed as well.
Consequently, a comprehensive and versatile standard shall be newly devised.\\
%where Resource Description Framework (RDF), Web Ontology Language (OWL), and SPARQL feature the architecture.\\
\indent As semantic media was initially geared towards web activities, the evolving standards in semantic media did not focus on defining scenes. 
In broadcasting verticals, \emph{TV-Anytime}\cite{TV-Anytime}, \emph{OMA BCAST} \cite{OMA BCAST}, and \emph{ATSC 3.0 A/332} \cite{A/332} emerged but were oriented to search and management.
\emph{MPEG-21} \cite{MPEG-21}, another major standard, delved into the distribution and transaction of multimedia items.
Digital rights management has been of its interest lately, particularly the technologies to ensure copyrights at a microscopic level. 
In this context, semantic features in media content, e.g., the features of objects appearing in video scenes, could be used in a broader sense, hence catalyzing the conceptions for semantic descriptions.
New conceptions of semantic descriptions would function as a gear in the collaborative system domain. 
A delivery system infused with semantic awareness or context sensitivity will elevate the network delivery ecosystem, ultimately moving towards semantic-native delivery$-$another future in intelligent X-casing.

\section{Transformation of Media Broadcasting: Generation at The End Device}
Our vision of GenAI-driven media casting (GMC) is described in this section.
The proposal in this section is roughly threefold: System concept, definition of service elements, and extension towards multi-prompt operations.
We first introduce the basis system design and continue with further extensions.
Following discussions detail the system and network aspects; related issues and challenges are also explored.

\subsection{System Design (Single-Source Operation)}
As mentioned, we assume receivers as generative, particularly to be capapble of composing media content from the received messages.
The generated media program is then directly presented on the screen if video service is considered\footnote{In fact, other forms of content can be considered. Audio, text, tactile stimuli, and other kinds of interface can be an option, seeking various immersiveness.}.\\
\indent The network serving this receiver can either be terrestrial broadcasting, cellular broadband, or wireless local access networks.
Any kind of digital radio technology can be employed\footnote{Wired ethernet and satellite connections, and the hybrid use of all mentioned technologies could also be counted.} if it is capable of transmitting the prompt data of interest and compatible with the semantic protocol thereof.
Nevertheless, the considered network differs from the conventional multimedia delivery, e.g., digital television (DTV) broadcasting and enhanced mobile broadband (eMBB), since it transmits only the prompt data instead of pixel-wise encoded video frames.\\
\begin{table}[!t]
\begin{center}
\caption{List of Terms}\label{terminology}
\begin{tabular}{|c|c|c|}
\hline
Class&Proposed GMC&Conventional\\
\hline\hline
Overall System&PDM system &Legacy media system \\
\hline
Serving Network&PSN &Conventional network  \\
\hline
Receiver&GRT &Non-generative terminal  \\
\hline
Transmit Data &Prompt package &Pixel-wise payload, \\
\hline
Service Retrieval &CG &Video decoder \\
\hline
\end{tabular}
\end{center}
\vspace{-0.5cm}
\end{table}
\indent Henceforth, we refer to this form of radio service network as \emph{prompt supply networks (PSN)} to tell it apart from conventional systems.
In the context of the end-to-end system operating upon this new type of mechanism, we use a term \emph{prompt-driven media (PDM)} system while the classical counterpart is denoted as \emph{non-generative} system.
Likewise, the considered receiver set is referred to as \emph{generative receive terminal (GRT)}, while specifying the GenAI functional system as a \emph{content generator (CG)}.
These definitions are summarized in Table \ref{terminology}.\\
\indent \textbf{Proposal 1 (Prompt-driven media systems):} The first part of our proposals consists of this idea of PDM systems.
As multi-prompt scenarios will later be addressed as well, we here use the term \emph{single-source operation ($\mathcal{O}_S$)} to specify the basis system actuated by a single prompt package.\\
\indent The \emph{prompt package} denotes a bundle of coordinated input messages (prompts) that constitutes a single intended service program.
Components in a prompt package are referred to as \emph{service elements}.
Such a package-wise definition is due to our broad-sense usage of the term prompt.
That is, this paper does not confine prompts to text instructions but allows them to encompass supplementary data across a wide range of modalities.
Either image, audio, layout, metadata, API, and program engines (graphics or physical) can form a prompt unit.\\
\indent Let us consider a \emph{VideoCrafter1} composing a video clip from a reference footage and a text guidance.
In this example, a text prompt and an image prompt constitute a prompt package.
For the sake of clarity, we define a unit of prompt package as dependent on the service provider's intent, i.e., a prompt package corresponds to a single intended service program.\\%\footnote{This may remind one of a concept of a complete delivery product (CDP). However, prompt packages differ from CDPs since CDPs represent the largest sets embracing every coordinated set. For instance, in a scalable video case, as an analogy, a CDP includes a base layer and every associated enhancement layer. By contrast, each valid combination of layers may correspond to a prompt package if it can be comprehended to a presentable video.}.\\
\begin{figure*}[!t]
\centering
\vspace{-0cm}
\includegraphics[width = 1.4\columnwidth]{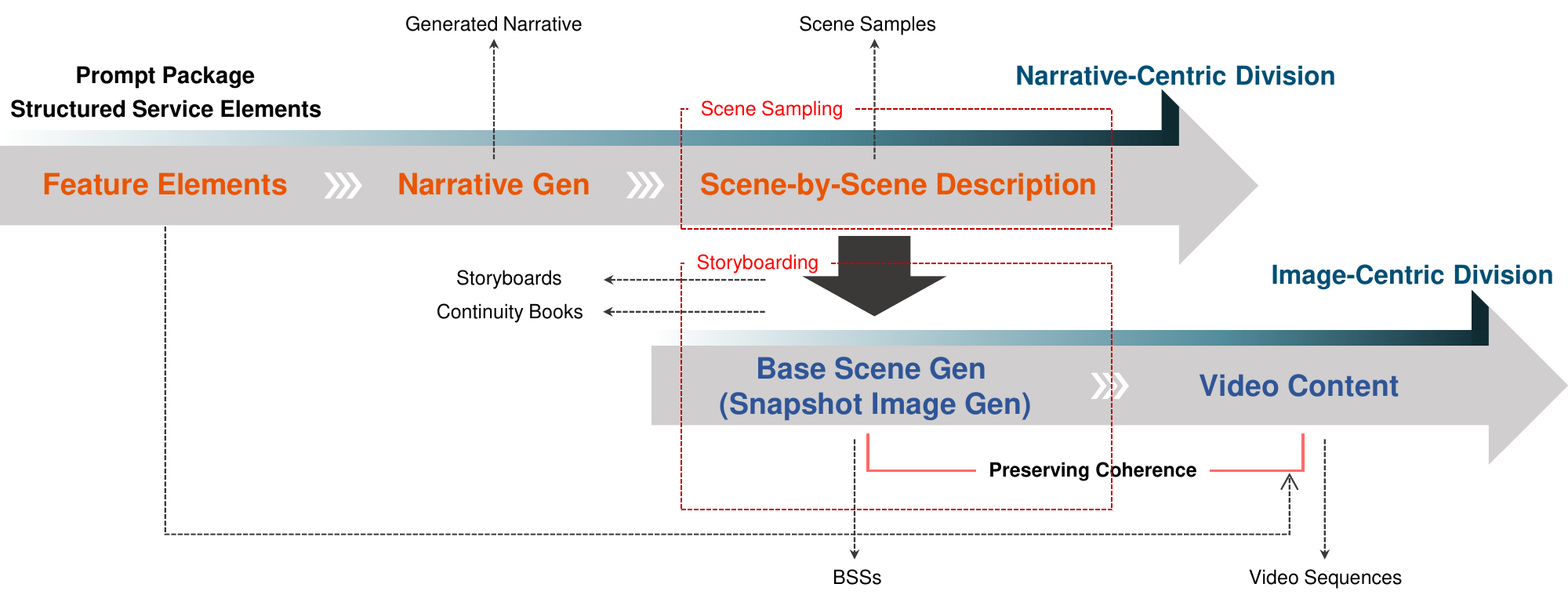}
\vspace{-0cm}
\caption{Example sketch of functional streamline in GRT.}\label{blockflow}
\vspace{-0.3cm}
\end{figure*}
\indent Regarding GRTs, the considered system is still in the early stages of development.
Nevertheless, the previous works have revealed the possibility and dim directions forward.
As a rough approach, we can na\"{i}vely scheme a cascaded assembly of generative modules.
Fig. \ref{blockflow} sketches one of such possibilities.\\
\indent This example splits the process into two parts: Narrative- and image-centric divisions.
The narrative-centric division builds the storyline and extracts storyboards therefrom.
Deduced from short pieces written by \emph{GPT-4} \cite{GPT-4}, further advanced language models will yield decent stories, possibly securing relevant coherence, consistency, and emotional resonance over a long plotline.
Although it was human-assisted, \cite{ReelFramer} partly demonstrated such potential by applying \emph{GPT-4} and \emph{MiniLM} \cite{Minilm} to scriptwriting and \emph{DALLE-2} to storyboarding.\\
\indent The storyboarding will proceed based on the scenes sampled from the generated narrative.
To this end, \emph{scene sampling} shall happen in the text domain first, preceding T2I transfer.
This scene sampling block crops essential cuts from the story, the representative scenes whereby connecting scene-by-scene will prevent contextual discontinuity in the film product.
Such \emph{scene samples} can be chosen event- or episode-wise and shall include pivotal scenes.
Naturally, denser sampling will enhance coherence and consistency, but will also escalate the computational burden as a consequence.\\
\indent The scene samples are then translated into the image domain, thereby producing storyboards.
Each storyboard can start with a rough sketch but is later enriched to form footage with proper resolution, specifically adequate to be fed into the I2V generator part.
We hereafter designate this refined snapshot footage by \emph{base scene snapshot (BSS)}.
Depending on the implementation, only the embedding data of BSS might be required instead of the actual footage embodiment.\\
\indent Each piece of storyboard (or BSS as well) can identify objects, background elements, their arrangement and characteristics, interactive access points, associated attributes, atmosphere styles, and other annotations.
Dynamic features like the movements of objects, changes in view angle, and adjustments to camera focus can be documented within the storyboard, or alternatively, we can detail them in a separate \emph{continuity book}.\\
\indent One continuity book is linked to one or more storyboards.
Note that each BSS is later stretched into a sequence of video frames.
A continuity book directs intended details to avoid \emph{oddity} within the continual sequence, especially motion and spatial deformation, or appearance and vanishment of features.
Basically, each continuity book is defined as interrelated to each storyboard: The $n$th book, {\fontfamily{cmtt}\selectfont continuity book$(n)$}, corresponds to {\fontfamily{cmtt}\selectfont storyboard$(n)$}.
The reference relation between {\fontfamily{cmtt}\selectfont continuity book$(n)$} and {\fontfamily{cmtt}\selectfont storyboard$(n)$} shall be indicated, where we can use the header and pointer tools or else.\\
\indent As mentioned, {\fontfamily{cmtt}\selectfont continuity book$(n)$} can be linked also to {\fontfamily{cmtt}\selectfont storyboard$(n\!\pm\!1)$} and others.
For example, a new object {\fontfamily{cmtt}\selectfont obj$_k^{n+1}$} may break into the frame in the next storyboard.
In that situation, the clip sequences created from the former storyboard shall include continuous movement of {\fontfamily{cmtt}\selectfont obj$_k^{n+1}$} entering in, which finally settles at the position described in the next storyboard.
Accordingly, for such particular features meant to be specified, it would be helpful to interlock the continuity books of interlinked storyboards by sharing some descriptions among each other,
e.g., identifying some common objects if needed.
Albeit the potential of GenAI might exhibit a smooth continuity without additional guidance, the continuity books would help expedite qualified productions.\\
\indent The continuity books and storyboards are generated at the same time, precisely originating in scene samples.
This procedure may reference the generated narrative again because the continuity books deal with differential aspects and transition.
The importance and role arrangement between continuity books and storyboards may depend on the density of scene sampling.\\
\indent The last part, converting BSSs into video frames, may harness the off-the-shelf I2V models.
Yet, for compatibility with the mixed-modal structure of storyboard and continuity book, and to support a wide range of modalities, the BSS-to-video composer should embrace further innovations.
In addition, the resolution of the video stream can be improved by appending consecutive upscaling/refinement/SR at the end part, likewise with \cite{Imagen_Video}, \cite{Singer}, and \cite{Zhang_XL}.\\
\indent \textbf{Remark 5 (Multi-modal contributions):} 
This design incorporates versatile learners whereby both input and output are mixed-modal.
To exemplify, imagine an advertisement scenario featuring the natural exposure of manufactured goods during an unfolding drama.
The GRT acquires the goods' snapshots and other information contained in the prompt package.
In this case, the generated narrative has to place or cite such information (e.g., illustrations) at proper positions in line.
The resultant scene samples, storyboards, and continuity books also include related information attached.\\
\begin{figure}[!t]
\centering
\vspace{-0.2cm}
\includegraphics[width = 0.95\columnwidth]{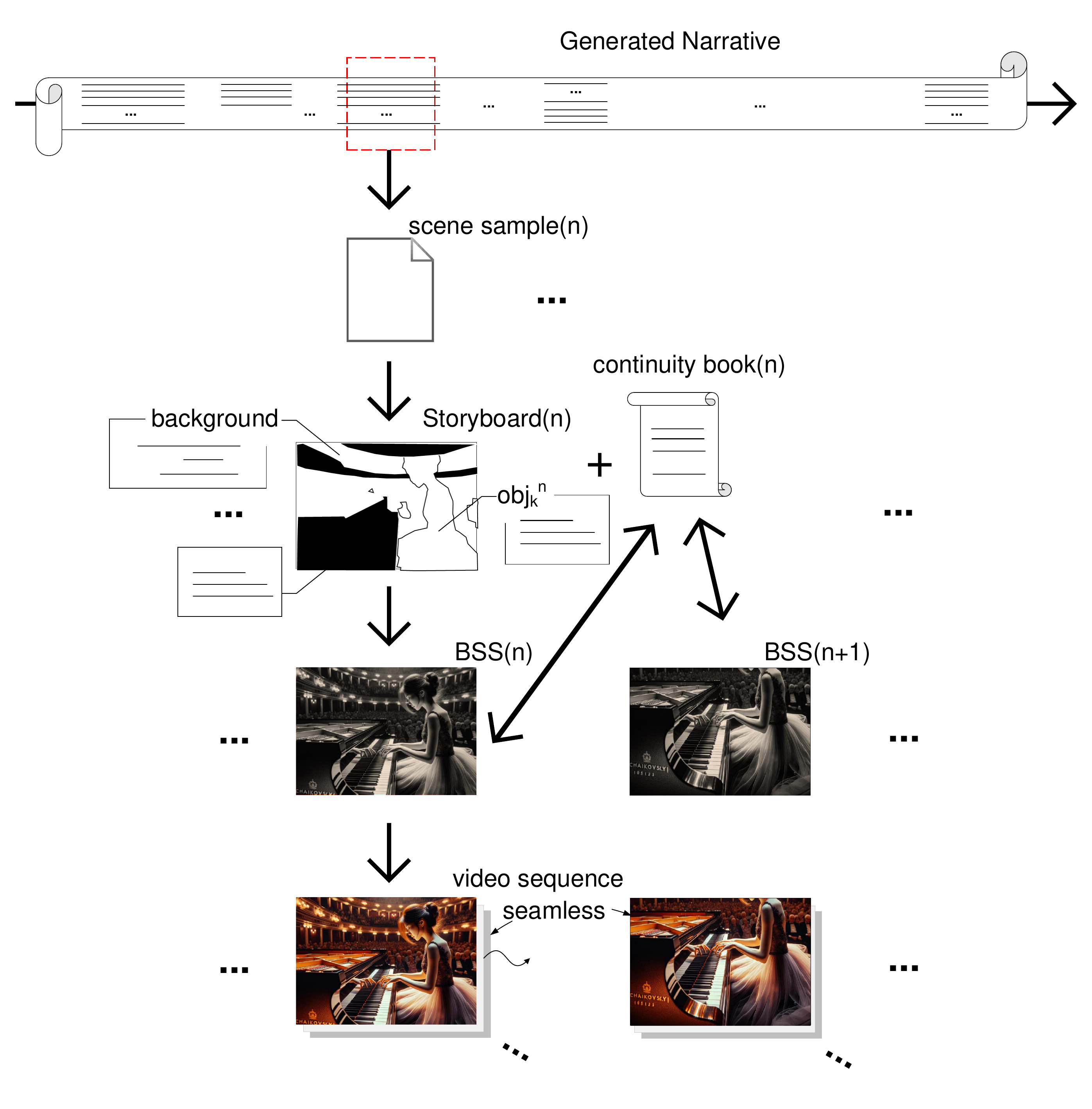}
\vspace{-0cm}
\caption{Schematics of CG workflow.}\label{systemex}
\vspace{-0.5cm}
\end{figure}
\indent Therefore, each block has to be capable of comprehending mixed-modal inputs and generating mixed-modal outputs as well.
Note that we also incorporate API and graphics/physics engines as possible forms of prompt elements.
For instance, the motion and mobility in video content may depend on such engines.
The trace of bounce movement$-$probably dependent on the surface properties and physical rules in the virtual world$-$, light effects, and many others can be engine-dependent.
The events, actions, and their descriptions in the generated narrative, and also their realization in the following stages$-$from storyboard to the final video sequence$-$could be conditioned by such service elements during each generation procedure.
Hence, the generated narrative, storyboard, continuity book, and BSS may invoke the related information.\\
\indent Moreover, mood and style could be handled as a class modality, and the system could utilize this information at each stage as well.
Plus, the media service program may sometimes offer user interactive functionality.
From the user interface (UI) standpoint, the access point for such interactive functions can be located in or overlaid on the moving pictures.
Accordingly, the location, form, and other attributes might need to be described within the intermediate outcomes during GRT operation.\\
\indent \textit{Challenge (In-context coherence):}
Maintaining context coherence is the key problem in this system. 
In this regard, we introduce several bonding devices.
Most importantly, the architecture in Fig. \ref{blockflow} itself is devised for this purpose, where the overall narrative is generated first and used as an underlying seed in the entire procedure. 
The continuity book is also one of the bonding devices, and we also establish channels through which every system stage can interact with semantic features obtained in preceding stages.
One could anticipate semantic consistency and spatiotemporal coherence to be seized accordingly.
The detailed implementation is an open problem but might be clued by previous studies, e.g., \emph{base stage} in \cite{Zhang_XL}.\\
\indent Our primary focus is on short-form videos and advertisement use cases, which makes coherence and continuity issues relatively manageable.
However, decent designs and innovations will top up time scalability and finally bring long-running films affordable.\hfill$\qed$\\
\indent \textit{Challenge (Time scheduling):} 
The PSN should take into account the time scheduling of media content, ensuring the avoidance of conflicts in time management among the intended services.
The prompt package could include information on runtime and provide timestamps.
In response to such information, the CG can timestamp the scene samples, storyboards, and BSSs, while the video frames could also be marked by representation time.
More intricate designs are possible, and Section III-C2 briefly revisits this aspect. \hfill$\qed$\\
\indent \textbf{Proposal 2 (Structured prompt definitions for PDM):}
Again, as first claimed in Section II-B, a structured grammar of prompts could propel GMC systems effectively.
We seek the detailed design of the prompt structures and the inclusion of rules within prompt packages to delineate the relationships among service elements.
These structures should be tailored for each modality while ensuring efficient cross-modal compatibility.\\
\indent Furthermore, for stable system operability across various GRTs, we aspire towards a unified framework that can achieve broad compatibility.
The volume of these projects is seemingly extensive, so it will necessitate collective contributions from the research communities and industry.
\subsubsection{Open problem, Design of prompt description standard}
These concerns instantly direct us to a standardized prompt design.
Note that a PSN would serve diverse GRTs fabricated by different manufacturers.
Simultaneously, each GRT may move access to different PSNs, and we can imagine a situation in which the service provider is changed in the next service session.
Thus, a wide-compatible framework is required to guarantee feasibility in practice. 
%Recent progress in media distribution has globally sought various kinds of compatibility across diverse components, particularly aiming for interactive and efficient services \cite{Ahn_5G}.
This paper suggests establishing a standardized protocol in this context, continuing the argument raised in Section II-B and II-C.\\
\indent The advantage of standard formatting is more than the above.
To be noticed, the CG models in GRTs should be first trained before putting into use.
For best performance, the prompts shall also be designed interactively, through mutual compromise with trained GRT models.
If the structure of prompts is optimized at the initial deployment stage, probably based on extensive experiments, the subsequent releases of GRT models only have to concentrate on training the CG upon the given, common, prompt structures.
Generally speaking, the release of new CG models would be more frequent than the revision in prompting standard.
Even if the prompt standard is revised, the CG model could be offline trained at the vendor testbed and then transferred to the GRT devices.\\
\indent Additionally, the advance in automatic prompt augmentation technologies would assist system fidelity \cite{Datta}.
Even if concise structures limit the prompt details, a prompt augmentation module will expand the prompt data, enabling the CG to acquire sufficient knowledge and generate a qualified output.
Such complement could be well-aligned with device-mounted SLMs \cite{phi-2}.\\
\indent At first, we can simply try applying semantic descriptions in the existing specifications like MPEG-7.
However, the existing frameworks defy flexible service.
For instance, MPEG-7 lacks abilities to specify the features within the picture domain, i.e., it is impossible to map the semantics to pixels accurately.
Accurate layout definition is not allowed as well.
To animate GMC beyond borderline feasibility, a comprehensive and versatile standard shall be newly devised, with the insights from Remark 5 guiding the design process.\\
\indent Additionally, the prompt structure can consider automated CoT for its design \cite{Shum}.
Some elements could be set as a sequence, preparing CoT, thereby seeking enhanced prompting.
The related context is previously briefed in Remark 4.\\
\sfdagger \textit{ Transmission of graph-based description}\\
\indent Once mentioned in Section II-C, graph-based description has been widely studied.
MPEG-7 Semantic Description defined a graph representation, as shown in Fig. \ref{xml}.
In the context of throughput efficiency, delivering information through a graph could be more efficient than using verbose sentences.
\cite{Wang} AA demonstrated this bit-saving opportunity, albeit in the context of knowledge graphs and text-oriented tasks. 
Therein, relational information was expressed in two-token sequences, effectively compressing textual sources into a small number of semantic tokens.
Thus, the design of the PSN prompt structure could consider the utilization of graphs as one of the possibilities. \hfill$\qed$
%\begin{align}
%Z(\mathcal{G}_i) = \sum_{g=1}^{G_i}(S_{i,j}^g+S_{i,k}^g+2)
%\end{align}

\subsubsection{Additional augmentation}
Despite its super-personalized trait, the described system mechanism is basically linear.
One may see the user experience in this GMC as passive, whereby the user participates little in formulating the content.\\
\indent In fact, the GMC service can accord some self-determination to users on top of $\mathcal{O}_S$.
Such an idea is achievable by appending add-on programs to the CG interface.
The flexibility of PDM should again be recognized, whereby only the substance for CG is received, which is editable at the GRT.
If the user installs a PDM-oriented app, this add-on program can append or edit the received prompt package according to user preferences.
This procedure is simply carried out before fetching the prompt package to the CG.\\
\indent The add-on program can screen censored or disliked features\footnote{Therefore, this can be understood as a protection program as well \cite{K.Liu}.}, or, on the contrary, replace the character's appearance with the preferred actor.
It is also possible to distort the storyline to revolve around the object of interest.
In this context, we can understand this system as one form of metaverse interacting with individual users.
Plus, one should notice that, it is not only a customized manipulation reflecting the users' bias but also applicable to the practice of enforcing regulation.\\
\indent \textbf{Example (Location-hooked service):}
One conceivable example is reflecting the real-world surrounding space in the generated media content, namely, a \emph{location-hooked service}.
Many street view services these days provide 360$^\circ$ online views for arbitrary locations, captured through on-site photography: e.g., \emph{Google Street View}\footnote{https://google.com/streetview}, \emph{Look Around}\footnote{Apple Maps application}, and \emph{Bing Streetside}\footnote{https://microsoft.com/en-us/maps/streetside}.
Knowing the current location, the GRT can access those street-view databases to import the images of surrounding sceneries.
The GRTs can then instruct CGs to generate video content having the user's actual surroundings as its backdrop.
Simple inpainting, swapping the background like \emph{Deepfake}, might work but the CG could also build a proper story from the narrative stage.
Such manipulation is achievable by merging the location-hooked data into the prompt package, as will be detailed in Section III-C.\\
\indent A location-hooked service offers an opportunity for augmented reality (AR).
The user can immerse themselves in a seamless experience that continues visual perception, crossing in and out of the screen.
Such experiences can extend into broader dimensions, with the GRT also capable of learning the surrounding audio and vocal context. \hfill$\qed$
\subsubsection{Capacity of semantic channel}
Note that the proposed PDM system is a certain form of SC.
Designing prompt packages and service elements falls under a sort of semantic encoding process introduced in \cite{Seo_TCCN}.
\cite{Seo_TCCN} and \cite{Seo_TCOM} defined semantics-native communication (SNC) as a model in which a speaker maps an idea entity into semantic concepts and subsequently into symbols, culminating in semantic representation (SR).
In this framework, the SNC with pre-determined encoder and decoder agents were denoted as \emph{System 1 SNC}, while calling \emph{System 2 SNC} a more intelligent system applying contextual reasoning-based adjustment to encoders and decoders. 
Although the PDM system with standardized prompt packages initially operates as a System 1 SNC, fine-designed prompts will bring it closer to the performance of System 2 SNC.
For System 1 and 2 respectively, \cite{Seo_TCCN} derived the bit-length ranges in semantic coding for System 1 and 2, indicating the necessary throughput in physical channels.
With well-designed prompts, the volume of prompt data will fall between the results for System 1 and 2 SNCs.
%Under System 1 semantic coding:
%\begin{align}\nonumber
%-\sum_{c\in\mathcal{C}}p_{X_c}(&\textrm{TRUE})\log_2 \frac{p_{X_c}(\textrm{TRUE})}{\sum_{c\in\mathcal{C}}p_{X_c}(\textrm{TRUE})} \leq L(\mathbf{t})\\ 
%&\leq \sum_{c\in\mathcal{C}}p_{X_c}(\textrm{TRUE})\bigg\lceil\! -\frac{p_{X_c}(\textrm{TRUE})}{\sum_{c\in\mathcal{C}}p_{X_c}(\textrm{TRUE})} \bigg\rceil,
%\end{align}
%where
%\begin{align}
%p_{X_c}(\textrm{TRUE}) = \sum_{a\in\mathcal{A}}p_{X_c|A}(\textrm{TRUE})p_A(a)
%\end{align}
%Under System 2 semantic coding (contextual reasoning):
%\begin{align}\nonumber
%-\sum_{k = 1}^K \sum_{c\in\mathcal{C}}p_C^k(c;\mathbf{t})&\log_2 p_C^k(c;\mathbf{t}) \leq L(\mathbf{t})\\ 
%&\leq \sum_{k = 1}^K \sum_{c\in\mathcal{C}}p_C^k(c;\mathbf{t})\Big\lceil\! -\log_2 p_C^k(c;\mathbf{t}) \Big\rceil,
%\end{align}
%where
%\begin{align}
%p_C^k(c;\mathbf{t}) = \sum_{a\in\mathcal{A}}p_{C|A}^k(c|a;\mathbf{t})p_{A|C}^{k-1}(a|c_{k-1};\mathbf{t})
%\end{align}
\subsubsection{Future Topics to Be Discussed}
We here introduce some items to be additionally discussed.
The related issues are as follows.\\
\sfdagger \textit{ Traffic feasibility in model-transferring case.}\\
\indent At times, the PDM systems may need to transfer CG models to GRTs.
Upgrades in PSN, fine-tuning between prompt packages and CG, and additional compatibility inclusions could trigger the model updating.
To gauge the data delivery volume, the following information can be referenced.
\emph{Imagen Video} consists of $11.6\!\times\!10^9$ parameters, translating to around 20 GB with a 16-bit floating-point system.
Despite the compactness of prompt packages, model data is substantial.
Note that H.265 video with $1280\times720$ pixels and a 24 fps frame rate amounts below 10 Mbps.
Model transfer should occur as seldomly as possible, and the efficiency shall be improved. \hfill$\qed$\\
\sfdagger \textit{ Computation offloading.}\\
\indent Depending on the computing capacity of the GRT, the CG can delegate computing tasks to cloud and edge servers.
Alternatively, the content can be generated within the network and then transmitted to the user.
For instance, the CG can be deployed on the edge server.
In this scenario, prompt packages are initially sent to the edge, and the edge-mounted CG promptly forwards the generated content.
In such cases, pixel-wise encoded video data or other intermediate data forms must traverse the physical radio channel between the network front-end and the user.
However, the PDM may remain efficient compared to legacy media systems, as it reduces the volume of pixels traveling among routers.
 Pixel traffic among routers can potentially congest the network, surpassing the volume associated with proximal cloud computing at the edge.. \hfill$\qed$\\
\sfdagger \textit{ Is PDM always the winner in network efficiency?}\\
\indent Note that some forms of service elements can outweigh the video data delivered in conventional systems.
API could be more bulky than encoded video frames.
Therefore, PSNs would not always promise lower channel traffic.
Nonetheless, PDM systems still create desirable opportunities for such traffic efficiency. \hfill$\qed$
\subsection{Multi-Source Operations: Motivation from Economics Aspect}
\begin{figure}[!t]
\centering
\vspace{-0.2cm}
\includegraphics[width = 0.9\columnwidth]{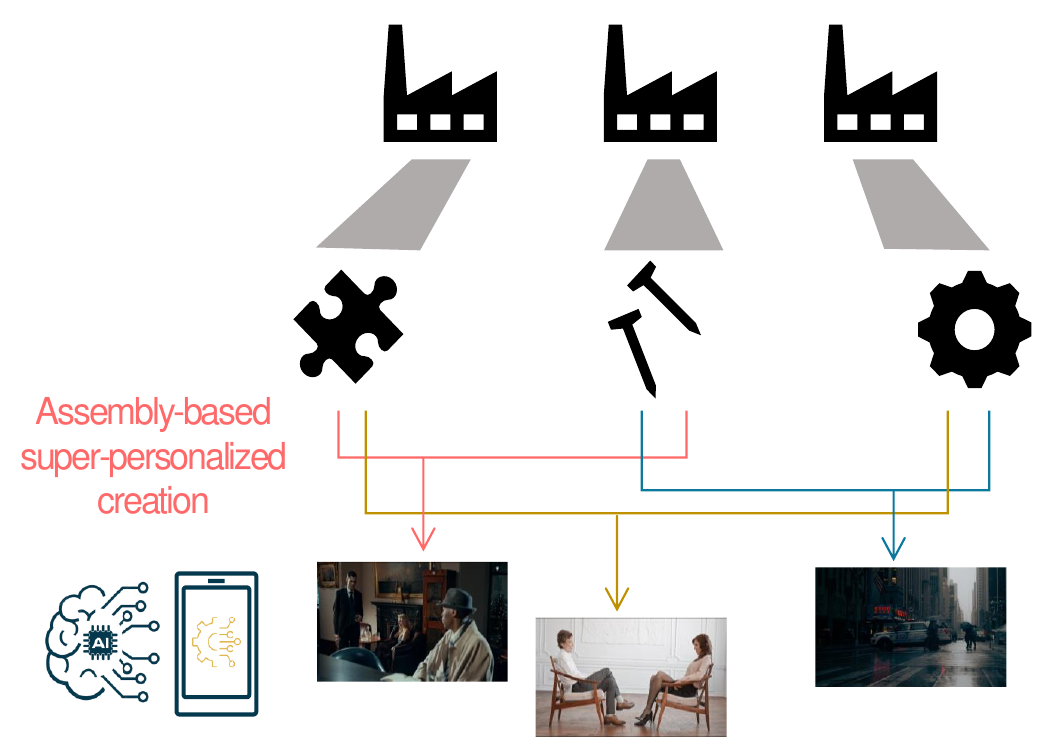}
\vspace{-0cm}
\caption{Conceptual illustration for $\mathcal{O}_M$.}\label{assembly}
\vspace{-0.3cm}
\end{figure}

\textbf{Proposal 3 (Multi-prompt operations):}
In addition to $\mathcal{O}_S$, this paper also proposes multi-prompt operations, which utilize multiple prompt packages at once.
These operations are alternatively referred to as \emph{multi-source operations ($\mathcal{O}_M$)} to name after $\mathcal{O}_S$.\\
\indent This concept is a renewal of the labor division, which contributed to the early success of traditional manufacturing. 
The potential of the supply chain division lies in the standardization of component specifications, enabling a multicast toward various different industries.
As a result, each manufacturing entity can concentrate its entire capacity on a partitioned microscopic process, which leads to economies of scale.\\
\indent This supply scheme is suitable for various end products, where the final output may differ depending on the combination and assembly. 
Our proposal is a variation of such division in the media domain, which can be schemed as Fig. \ref{assembly}.
\subsection{Multi-Source Operations: Description}
This paper introduces two classes of $\mathcal{O}_M$: Synchronous $\mathcal{M}_1$ and asynchronous $\mathcal{M}_2$.
Each concept is in parallel described in the subsections below.
$\mathcal{M}_1$ and $\mathcal{M}_2$ are first devised for different target environments, but fusion use can be envisaged also.

\subsubsection{$\mathcal{M}_1$ (Multi-prompt scenario 1, Synchronous)}
$\mathcal{M}_1$ considers a synchronous configuration in which multiple prompt packages are received at the same time.
Conceptually, this idea succeeds the \emph{add-on} augmentation in Section III-A2.
In effect, $\mathcal{M}_1$ may employ multiple prompt packages more aggressively than the previous augmentation notion.
Whereas the \emph{add-on} programs are basically built as subordinate to the main service, just customizing it, $\mathcal{M}_1$ embraces different service provider's prompt packages as equivalent input information.\\
\indent There are several possible ways to reconcile independent prompt packages, each of which can evoke independent cinema.
In essence, the underlying idea is to unify the prompt packages of interest into one integrated prompt package. \\
\indent First, we can begin with the case in which none of the prompt packages micromanages the storyline or direction.
One can consider two independent advertising prompt packages, which contain detailed information about the advertised products' appearance but limited elaboration on other directorial elements.
The GRT can then consider the images of those two advertised products as service elements within a single prompt package, hence creating a drama that presents both products within the mise-en-sc\`{e}ne.\\
\indent As an extreme case, suppose two independent prompt packages $\mathcal{P}_1$ and $\mathcal{P}_2$, each containing a piece of picture only. 
Let these pictures denoted by $\mathbf{p}(\mathcal{P}_1)$ and $\mathbf{p}(\mathcal{P}_2)$, respectively.
Suppose that the resultant media content of each would display a freeze-frame of the picture during the runtime when it is fed into the CG alone.
Then, simply put, an image manipulation steering the embedding vector over time$-$from that of $\mathbf{p}(\mathcal{P}_1)$ to $\mathbf{p}(\mathcal{P}_2)$$-$can form the media content generated by $\mathcal{M}_1$ using $\mathcal{P}_1\cup\mathcal{P}_2$ (see \cite{Ramesh1} for the image manipulation).\\
\indent Comprehending multiple prompt packages as one merged package, a \emph{merge-and-read} strategy in short, will work for further complex cases as well.
This strategy is actually identical to some $\mathcal{O}_S$ tackling a complex or flimsily planned prompt package.
The \emph{merged} prompt package will do work if there is no conflict among the different prompt packages therein.
Even if so, we can deal with it by assigning priority among the service elements.
Every service provider, i.e., prompt designer, may set element priorities, i.e., which service element shall be prioritized over others.
The prompt merging module for $\mathcal{O}_M$ could then sift the service elements at the merging stage, selecting top-$K$ for each package for example.
This measure allows the GRT to maintain complexity as manageable and avoid potential conflicts.
Accordingly, the definition of service element priority should be included in the prompt structure standard.\\
\indent On top of this setup, the GRT could also arbitrarily determine the main prompt, the primary contributor to composing the narrative.
The other prompt packages or elements are then set as sub-prompts.
Such a process can refer to the service element priority as well.

\subsubsection{$\mathcal{M}_2$ (Multi-prompt scenario 2, Asynchronous)}
On the other hand, $\mathcal{M}_2$ deals with a sequence of prompt packages arriving over time.
Precisely, a prompt package $\mathcal{P}_2$ could arrive in the middle of the $\mathcal{O}_S$ process for $\mathcal{P}_1$.
Let us denote by $\mathcal{C}_\Omega(\mathcal{P})$ the generated content prompted by a prompt package $\mathcal{P}$ and CG operation $\Omega\in\{\mathcal{O}_S,\mathcal{M}_1,\mathcal{M}_2\}$.
If the GRT decides to combine the services by means of $\mathcal{M}_2$, instead of postponing the presentation of $\mathcal{C}_{\mathcal{O}_S}(\mathcal{P}_2)$ until $\mathcal{C}_{\mathcal{O}_S}(\mathcal{P}_1)$ ends, the CG could seamlessly blend $\mathcal{P}_2$ with the remaining part of $\mathcal{C}_{\mathcal{O}_S}(\mathcal{P}_1)$.\\
\indent To this end, we come up with a segmented time schedule of $\mathcal{C}_{\mathcal{O}_S}(\mathcal{P}_1)$.
Let us first assume zero-processing time for content generation.
By monitoring the presentation of $\mathcal{C}_{\mathcal{O}_S}(\mathcal{P}_1)$, the GRT can backtrack the BSS, storyboard, and scene sample to which the current on-screen video cut belongs.
Then, the CG can apply $\mathcal{M}_1$ to $\tilde{\mathcal{P}}^c_1\cup\mathcal{P}_2$, where $\tilde{\mathcal{P}}^c_1$ indicates the parts of generated narrative, storyboards, BSSs, and related elements in $\mathcal{P}_1$ that correspond to the remaining part of $\mathcal{C}_{\mathcal{O}_S}(\mathcal{P}_1)$.
Clearly, the following video sequences are resulted from $\mathcal{C}_{\mathcal{M}_1}(\tilde{\mathcal{P}}^c_1\cup\mathcal{P}_2)$.\\
\indent Note that the time schedules of $\mathcal{P}_1$ and $\mathcal{P}_2$ would be different.
Only a part is likely to overlap between their runtime plans.
Suppose that $\mathcal{P}_1$ and $\mathcal{P}_2$ are scheduled in $[t_0,t_0+2\Delta T)$ and $[t_0+\Delta T,t_0+3\Delta T)$, respectively.
In this case, $\tilde{\mathcal{P}}^c_1\cup\mathcal{P}_2$ lasts in $[t_0+\Delta T,t_0+2\Delta T)$, and the rest following are devoted to $\mathcal{P}_2$ solely.
This point necessitates a more careful approach in terms of time scheduling.\\
\indent We consider the BSSs to be timestamped, identifying the time schedule of the resultant video frames.
Each storyboard and scene sample is coupled with BSS, so we can find the segmented time schedule based on the storyboard unit.
Otherwise, more intricate terms are also possible.\\
\indent Let us assume $[t_0+\Delta T,t_0+2\Delta T)$ as equivalent to the time segments $\{\tau_{n_i},\cdots,\tau_{n_f}\}$, which correspond to the scene samples $\{S_k^1,\cdots,S_{K1}^1\}$ and narrative flag points $\{N_l^1,\cdots,N_{L1}^1\}$ in $\mathcal{C}_{\mathcal{O}_S}(\mathcal{P}_1)$, respectively.
$\tilde{\mathcal{P}}^c_1$ consists of $\{S_k^1,\cdots,S_{K1}^1\}$, $\{N_l^1,\cdots,N_{L1}^1\}$, and related service elements.
One should here notice that $[t_0+\Delta T,t_0+2\Delta T)$ is primarily devoted to $\mathcal{P}_1$.
Hence, in $\tilde{\mathcal{P}}^c_1\cup\mathcal{P}_2$, $\tilde{\mathcal{P}}^c_1$ is taken as the main prompt until $t_0+2\Delta T$, whereas $\mathcal{P}_2$ is determined as a sub-prompt.
For the story unfolding since then, i.e., $[t_0+2\Delta T,t_0+3\Delta T)$, the CG elevates $\mathcal{P}_2$ to be the main prompt.
If not necessary, the features originated from $\mathcal{P}_1$ fade out in $[t_0+2\Delta T,t_0+3\Delta T)$.
For example, if $\mathcal{P}_1$ is an advertisement of a goods $G_1$ and $\mathcal{P}_2$ is for $G_2$, (i) $\mathcal{C}_{\mathcal{O}_S}(\mathcal{P}_1)$ in $[t_0,t_0+\Delta T)$ will display $G_1$ only, (ii) $\mathcal{C}_{\mathcal{M}_1}(\tilde{\mathcal{P}}^c_1\cup\mathcal{P}_2)$ in $[t_0+\Delta T,t_0+2\Delta T)$ will come along with both $G_1$ and $G_2$, and (iii) $G_1$ will vanish since $t_0+2\Delta T$.\\
\indent Situations otherwise could be tackled by applying such principles.
We can also generalize this procedure with arbitrary $N$ prompt packages $\mathcal{P}_2,\cdots,\mathcal{P}_{N+1}$ arriving in the middle of the $\mathcal{O}_S$ process for $\mathcal{P}_1$.

\subsubsection{Dynamics Sparked by $\mathcal{O}_M$}
\begin{figure}[!t]
\centering
\vspace{-0cm}
\includegraphics[width = 0.85\columnwidth]{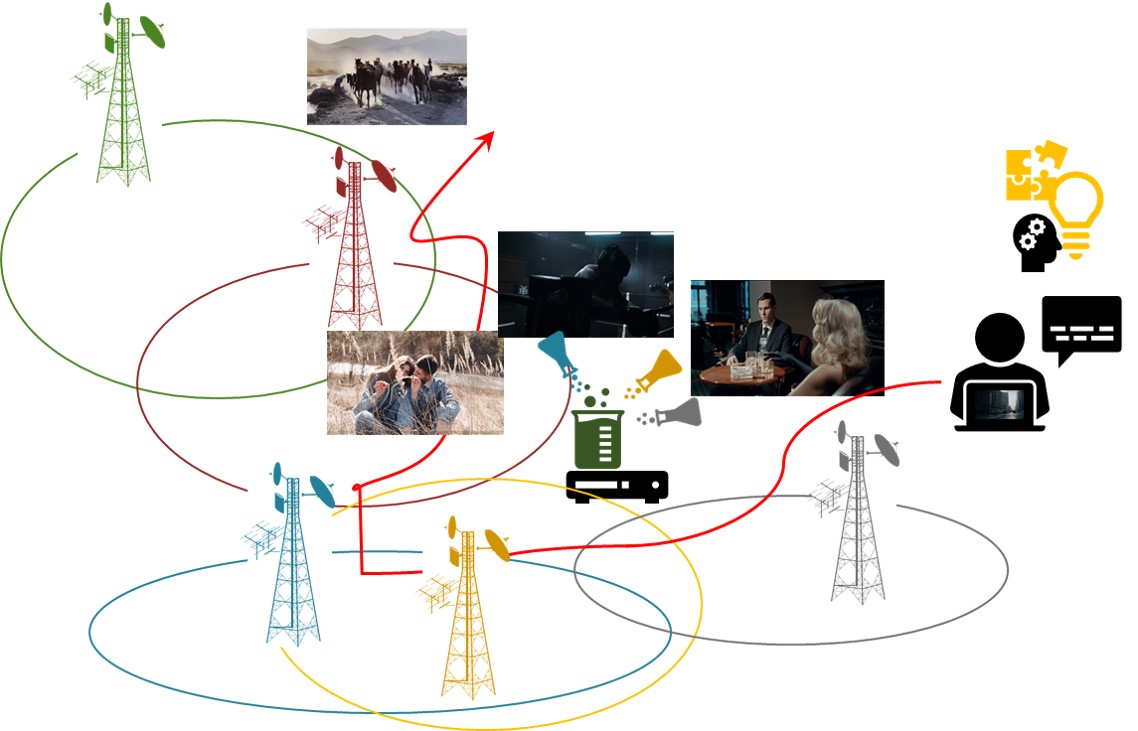}
\vspace{-0cm}
\caption{Dynamic transition of narratives in a moving GRT scenario under $\mathcal{O}_M$.} \label{diagram}
\vspace{-0.3cm}
\end{figure}

By integrating a combinatory mechanism, $\mathcal{O}_M$ deepens the dynamism in content creation.
This implies that different assemblies bring about varied moods of content, as depicted in Fig. \ref{diagram}.\\
\indent For an example with physical coverage segmentation, consider a scenario with six service providers each transmitting a distinct prompt package:
denoted as $A$, $B$, $C$, $D$, $E$, and $F$.
Assuming these providers act as broadcasters, some of their cell coverages overlap, while others do not.
We assume that the GRT can receive multiple service providers' signals simultaneously if the GRT is within the coverage.
Suppose a situation where a mobile GRT traverses through a zone ${A, B, C}$ are available, then where with ${C, D}$, and finally detects ${A, E, F}$.
With $\mathcal{M}_2$ and $\mathcal{M}_1$ combined, the CG will generate a drama initially characterized by $A \cup B \cup C$, transitioning to incorporate atmospherics from $C \cup D$, and evolving into a new plot with the features of $A \cup E \cup F$. 
The generated cinema could seamlessly transition from noir to action and comedy while maintaining consistent objects or characters.\\
\indent This thought suggests extremely dynamic dramas that can be made by localized assembly of service data.
Such examples are not limited to the physical coverage segmentation concept, but also can be applied to virtual dimensions, such as social classification (see the targeted ad example in Section IV).

%In this concept, each broadcaster's service will consist of service elements and prompts. 
%Users will not consume the same finished media content but rather unique media content based on their individual situations.\\
%\indent We expect that in the near future, dynamic media generation systems like this will be more suitable for shorter-length media forms, such as modern YouTube shorts and TikTok videos, rather than content with long-term plots. 
%It is believed that this transition aligns with the current media ecosystem, which is rapidly shifting towards very short-length videos. 
%Additionally, it is considered suitable for applications like advertising.

\section{Use Cases}
This section lists some use cases for the proposed GMC.
Basically, the GMC features advantages in two parts: ($i$) Fatigue-avoiding content variation and ($ii$) Compressed delivery against capacity-limited channels, aiming at high-performance entities.
The following examples illustrate the benefits respectively, where the $1^\textrm{st}$ to $3^\textrm{rd}$ items belong to ($i$) and the $4^\textrm{th}$ to $5^\textrm{th}$ belong to ($ii$).\\
\noindent \sfdagger \textit{ Advertising Screens in or on Mass Transportation}\\
\indent The GMC is particularly advantageous when handling content forms prone to fatigue, such as advertisements.
Most of the advertising screens in mass transportation today display repetitive sequences with limited content.
Some local servers for such displays, mounted in the vehicles, could be connected to the internet, but in many cases, they exploit pre-fetched files in the local storage.\\
\indent The GMC could keep refreshing the presented content without loading new files to the local server.
If the GRT acquires several prompt packages, offline or online with a small amount of delivery, the passengers can watch new advertisements every time.
Moreover, location-hooked service can be applied, which will offer an immersive impact to the viewers in moving vehicles.
A similar can be applied to the screens attached to the vehicle exterior, targeting outdoor pedestrians. \\
\noindent \sfdagger \textit{ Digital Signage}\\
\indent The GMC-aided digital signage is another version of the aforementioned example.
Unlike the motorized model above, this example assumes stationary displays.
As the device is fixed in place, digital signage could better lay on location-hooked service and other additional augmentation.\\
\noindent \sfdagger \textit{ Pinpoint-Targeted Advertisement}\\
\indent The proposed $\mathcal{O}_M$ and customized augmentation present opportunities of precise targeted advertising.
Traditional targeted ads are constrained by a predefined set of \emph{prepared} ad programs, limiting their adaptability \emph{resolution} to individual user attributes.
The cardinality of this set is coupled with the classification of target users.\\
\indent Suppose that there is a set of attributes $\{a,b,c,d,e,f\}$ expected among general users.
The ad provider may prepare content for $\{a,b,c\}$, $\{a,c,e\}$, and $\{b,e,f\}$.
In fact, some users may have $\{a,b,d\}$ and highly against $c$.
Nonetheless, the ad for $\{a,b,c\}$ will be delivered to these users under the rule of majority.
This is an inevitable off-target failure due to the limited requirement of prepared content.\\
\indent In contrast, GMC and $\mathcal{O}_M$ overcome this problem.
If the prompt packages are prepared for every attribute, each prompt package can be independently guided to apt users.
The pinpoint-targeted ad will then be automatically generated based on the combination of \emph{arrived} prompt packages.
This aspect will highly improve the target ad strategies.\\
\noindent \sfdagger \textit{ Qualified Videos in High-Speed Mobile Environments}\\
\indent High-speed mobile environment has been a challenge to wireless networks.
For terrestrial broadcasting conveying high-quality videos, mobile reliability issues are problematic \cite{Ahn_Div}.
In high-speed transportation scenarios, local storage is often employed to deliver cinematic entertainment services to passengers. 
The PDM system could effectively address this issue if the capability of CG is sufficient.
This mobile environment precisely falls within the realm of capacity-limited channels accommodating high-performance devices.
Matured PDM systems would deliver video entertainment by accessing compact prompt packages exclusively. \\
\noindent \sfdagger \textit{ Space Communications}\\
\indent With growing concerns about space development, space communications have gathered noticeable interest.
However, the deep space environment, characterized by cosmic propagation distances, poses challenges related to channel attenuation.
Therefore, the PDM system emerges as a potential solution for media delivery in scenarios where precise copy accuracy is not paramount.

\section{Concluding Remarks: Futures of Multicast/ Broadcast Networks}
As a part of \emph{IEEE Trans. Broadcast. Special Issue on Intelligent Multicast/Broadcast Services over 5G/6G}, this paper introduced a GenAI-driven blueprint for upcoming transformation in multimedia casting.
Still, the contribution of the proposed scheme will reach a portion, where the evolutions in multicast/broadcast technology are anticipated to occur across diverse disciplines.
Most notable prospects expected in the broadcast domain include the following items.
\subsection{Dynamic Coordination upon Broadcast-Broadband Convergence (BcBbC)}
\indent The broadcast networks are evolving into an interactive form.
The IP-native development of ATSC 3.0 catalyzed such movements by enabling flexible interplay with broadband networks.
Additionally, the endeavors on 3GPP-native multicast/broadcast technologies, such as \emph{FeMBMS} and \emph{NR MBS}, encouraged the discussion more \cite{SKAhn_5GB}.\\
\indent The BcBbC can take place in various layers, from the application layer down to the physical layer.
Mixed-mode transmissions in the physical layer will escalate achievable throughput \cite{Ahn_NOMA}, \cite{Ahn_CRAN}; offloading and handover across the transport to application layers will boost the network efficiency and support seamless experience \cite{Ahn_BCBB}; broadband-aided feedback will enrich user interactivity within service content \cite{Ahn_5G}.\\
\indent Specifically, we imagine a dynamic program scheduling for broadcast channels.
On top of flexible BcBbC, the broadcast networks could adaptively transmit the streaming content temporarily popular, e.g., live streaming programs popping up in an ad-hoc sense.
An intelligent load-splitting of online game data streams between unicast and broadcast is also in the scope.
Further various applications like BcBbC-collaborative security can be addressed.
\subsection{Enabling Technology for Infrastructure in Other Domains}
\indent The \emph{convergence} of broadcast networks will expand toward further various domains.
Preserving the public-engaged nature, broadcast networks can become a foundational infrastructure supporting critical industries.\\
\indent For example, there were efforts to replace backhaul links with terrestrial broadcast signals \cite{Hunter_SI}-\!\!\cite{Li_ITCN}.
With excellent stability, broadcast transmissions can co-locate the public media program and inter-tower backhaul signals within the same physical layer frame, thereby amplifying spectral efficiency \cite{Kang_MIMO}.
The occupied frequency band, which is essentially granted for public-oriented use, will consult the public interests in a broader sense.\\
\indent More ideas will make broadcast networks further versatile.
The broadcast signals could be used to promise standard time synchronization in critical industries, e.g., finance and power transportation, even when the Global Positioning System satellite signals are lost.
\cite{NAB}-\!\!\cite{NAB3} proposed ideas to establish such a backup network based on ATSC 3.0 on-air networks, so-called broadcast positioning system.
Moreover, broadcast networks may support future, automated transportation systems by distributing public intelligent traffic system (ITS) control data.
The traffic efficiency will keep enlightening application opportunities, such as wide-region software updating in the era of immersive internet and autonomous vehicles.
\subsection{Supplier for GenAI-Driven Media}
\indent This GMC item is to which this paper was devoted.
Inspired by the deployment of generative models in user devices, we visualized a new media ecosystem that can emerge accordingly.
Our proposal suggested a transformation of multicast/broadcast and unicast networks into PSNs.
According to this design, media broadcasting will depart from distributing the media content already finished at the production house.
Instead, the media content will be generated at the user side, GRT itself or GRT-involved cloud computing, while the service providers will supply prompt packages.\\
\indent To this end, this paper designed CG operations $\mathcal{O}_S$ and $\mathcal{O}_M$ facilitating the proposed PDM system.
The users will then be allowed to experience fresh and super-personalized content every time.
Moreover, since the PSN delivers lightweight prompts rather than exhaustive pixel-wise encoded video streams, this idea will seize significant network efficiency as well.
To stabilize the system and promise fine average quality, the industry could employ structured formats for prompts.
From this point, we suggested designing a structured standard for prompt packages, also envisaging a broad range of multi-modality within prompt packages.\\
\indent The GMC uniquely reveals a novel opportunity to promise a user-specific experience within linear delivery.
This advantage will particularly shine in content forms prone to fatigue, such as advertisements.
As conceived in this paper through customized augmentation, location-hooked service, and $\mathcal{O}_M$, the GMC's personalized and fresh characteristics will be further magnified.
Resultantly, future multicast/broadcast services will achieve substantial dynamics unseen in the previous systems.

\ifCLASSOPTIONcaptionsoff
  \newpage
\fi


%\end{IEEEbiography}


%\newpage



%\vfill


\end{document}